# Statistical theory of the excited strip domain structure

**E.S. Denisova**

Physical and Technical Department, Sumy State University, 2 Rimskiy-Korsakov St., 244007 Sumy, Ukraine

**Abstract**

A statistical theory of the strip domain structure excited in a bubble film by an oscillating magnetic field is developed. The theory is based on the consideration of the strip domain structure as a thermodynamic system characterized by the spectrum of domain walls oscillation and an effective temperature that is caused by an oscillating magnetic field and film nonuniformities. We found the thermodynamic characteristics of that domain structure and calculated its period as a function of the frequency and amplitude of an oscillating magnetic field.

*PACS:* 75.70.Kw, 05.70.Ce

*Keywords:* Magnetic film; Domain structure; Thermodynamic functions; Equation of state

# 1. Introduction

The period change effect of the strip domain structure (SDS) induced by an oscillating magnetic field was observed first in silicon-iron plates [1,2]. Later it was also discovered in samples of yttrium-iron garnet [3], iron garnets [4], iron borate [5]. According to [6] the main peculiarity of that effect is a nonmonotonic dependence of the period $p$ on the frequency $w$ and the amplitude $H$ of an oscillating field. For describing the period change effect a few theoretical schemes were proposed. In [1] the equilibrium period was found from the condition of the minimum of the dissipated energy in a domain structure. Another approach to solving this problem based on averaging of the Landau-Lifshitz equation (over the field oscillations), and finding of the effective energy of domains and domain walls (DWs) was given in [7]. But the analysis carried in [6] showed that neither dissipative nor dynamic mechanisms describe even qualitatively the peculiarity noted above. That peculiarity is not described in the frames of model [5] either which is based on the supposition of a quasi-static character of the oscillating field influence on SDS.

In the present paper we develop a statistical theory of the excited SDS which qualitatively explains all peculiarities of the period change effect.

# 2. Effective temperature of the excited SDS

According to [8, 9] the spectrum of DW oscillation in SDS of a bubble film (see Fig. 1) consists of the acoustic branch $w_1(k)$ and the optical one $w_2(k)$ ($k$ is the dimensionless wave number). In perfect films an oscillating magnetic field $H\cos(wt)$ oriented along the axis $z$ (the axis $z$ of the Cartesian coordinate system is parallel to the easy axis of magnetization, the plane $xy$ is the central film plane, and the axis $y$ is oriented along the strip domains) causes only optical oscillations of DWs with $k = 0$:

$$m\ddot{x}_n + l\dot{x}_n + mw_2^2(0)x_n = (-1)^n 2MH\cos(wt). \qquad (1)$$

Here $x_n$ is the deviation of the $n$-th $(n = 0, \pm 1,...)$ DW from the equilibrium state, $m$ and $l$ are the mass and the damping coefficient of a DW unit area, respectively, $M = |\mathbf{M}|$ is the magnetization,

$$w_2^2(0) = 2\Omega^2 \ln\left(1 + \frac{\sinh^2(h/2)}{\sin^2(r/2)}\right), \qquad (2)$$

$\Omega^2 = 8M^2/mh$, $h = 2ph/p$, $r = 2pa/p$, $h$ is the film thickness, and $a$ is the width of those strip domains which are magnetized antiparallel to the axis $z$. In equilibrium the values $h$ and $r$ satisfy the equations $\P W/\P h = \P W/\P r = 0$, where

$$W = 4M^2 hL_x L_y \left[ h\frac{l}{h} + r\left(\frac{H_0}{4pM} - 1\right) + \frac{r^2}{2p} + \frac{4}{ph} \sum_{n=1}^{\infty} \frac{1-e^{-hn}}{n^3} \sin^2\frac{rn}{2} \right] \qquad (3)$$

is the SDS magnetic energy [10], $hL_x L_y (\to \infty)$ is the volume of the bubble film, $H_0 = |\mathbf{H}_0|$ is the external magnetic field, and $l$ is the characteristic length of the material.

In nonperfect films DWs also perform random oscillations in consequence of their interaction with the film nonuniformities. Therefore, we can consider the excited SDS as a thermodynamic system in this case and characterize it by an effective temperature. Assuming that the DWs are not bent under interaction and using the law of equipartition of energy [11], the effective temperature can be defined as $T_{eff} = mhL_y \langle \dot{X}_n^2 \rangle / k_B$, where $X_n$ ($\langle X_n \rangle = 0$) is the random deviation of the $n$-th DW from the equilibrium state, $\langle \rangle$ denotes an averaging over realizations of $X_n$, and $k_B$ is the Boltzmann constant. In the general case for calculating $\langle \dot{X}_n^2 \rangle$ it is necessary to have detailed information about the interaction of DWs with film nonuniformities. But if the correlation radius $r_c$ and the mean square fluctuation $H_f$ of the random field, which models the influence of film nonuniformities, do not exceed $\max|x_n|$ and $H$ on the order of magnitude, respectively, then we can estimate $\langle \dot{X}_n^2 \rangle$ as $\overline{\dot{x}_n^2}$ (the bar denotes an averaging on the period $2\pi/\omega$ of the oscillating magnetic field). In this case, using Eq. 1, we obtain

$$T_{eff} = \frac{(Hh)^2 L_y}{4k_B} \frac{(\Omega \omega)^2}{[\omega^2 - \omega_2^2(0)]^2 + (\omega \lambda / m)^2}. \tag{4}$$

Note that in our approach a concrete mechanism leading to the thermalization of SDS and permitting to consider it as a thermodynamic system is not essential; the main thing is that such mechanism exists.

Let us evaluate $T_{eff}$ for $\omega = \omega_2(0)$. If $H = 10$ Oe, $h = 10^{-3}$ cm, $L_y = 10^{-2}$ cm, and $\lambda / m\Omega = 5$, we have $T_{eff} \approx 10^8$ K, i.e. the effective temperature is much large than the thermodynamic temperature. This means that the entropy term $(T_{eff} S)$ in the free energy $F = W - T_{eff} S$ of the excited SDS can be predominant. In turn this means that the values $h$ and $r$ must be found from the minimality condition of $F$.

## 3. Thermodynamic functions of the excited SDS

Under suppositions made above, the thermodynamic properties of the excited SDS are the same as the thermodynamic properties of the classic one-dimensional crystal which has the temperature $T_{eff}$ and the Hamiltonian function

$$H(q, p) = \frac{1}{2} \sum_{j=1}^{2} \sum_k \left[ \frac{p_j^2(k)}{mhL_y} + q_j^2(k) \omega_j^2(k) mhL_y \right]. \tag{5}$$

Here, $k = r/N$, $r = 0, \pm 1, \ldots, \pm [N/2]$ ($[N/2]$ is the integer part of $N/2$), $N = [L_x/p]$ (for convenience $N$ is chosen as an odd number), and $q_j(k)$ and $p_j(k)$ are generalized coordinates and impulses, respectively. Using Eq. (5) and excluding from consideration the displacement of SDS as a whole, its statistical integral defined as [11]

$$Z = \frac{1}{(2\pi\hbar)^{2N}} \int e^{-\frac{H(q,p)}{k_B T_{eff}}} \prod_{j=1}^{2} \prod_k dq_j(k) \, dp_j(k) \tag{6}$$

($\hbar$ is the Planck constant) can be represented in the form

$$Z = \left( \frac{k_B T_{eff}}{\hbar} \right)^{2N-1} \frac{1}{\omega_2(0)} \prod_{k>0} \frac{1}{\omega_1^2(k) \omega_2^2(k)}. \tag{7}$$

To simplify the calculation of $Z$, let us consider the particular case when $H_0 = 0$. In this case $r = p$,

$$w_2(0) = 2\Omega \ln^{1/2}(\cosh(h/2)), \tag{8}$$

$$W = 4 M^2 h L_x L_y \left( h\frac{l}{h} + \frac{4}{ph}\sum_{n=1}^{\infty}\frac{1-e^{-h(2n-1)}}{(2n-1)^3}\right), \tag{9}$$

and the functions $w_1(k)$ and $w_2(k)$ can be approximated in the following way

$$w_1(k) = w_2(0)k, \quad w_2(k) = w_2(0)(1-k). \tag{10}$$

Using Eq. (10) to find $Z$, we obtain for $N \gg 1$ $Z = \left(ek_B T_{eff}/\hbar w_2(0)\right)^{2N}$. Hence, the entropy $S = k_B \ln Z$, the free energy $F = W - k_B T_{eff} \ln Z$, and the chemical potential $m = (\partial F/\partial N)_{T_{eff}}$ of the excited SDS take the form

$$S = 2N k_B \ln\left(\frac{e^2 k_B T_{eff}}{\hbar w_2(0)}\right), \tag{11}$$

$$F = W - 2N k_B T_{eff} \ln\left(\frac{ek_B T_{eff}}{\hbar w_2(0)}\right), \tag{12}$$

$$m = \frac{2ph}{L_x}\frac{dW}{dh} - 2k_B T_{eff} \ln\left(\frac{ek_B T_{eff}}{\hbar w_2(0)}\right) + 2hk_B T_{eff}\frac{\Omega^2}{w_2^2(0)}\tanh\left(\frac{h}{2}\right). \tag{13}$$

## 4. Equilibrium period of the excited SDS

Since $dF = -S dT_{eff} + m dN$ and the effective temperature $T_{eff}$ depends on $N$, the equilibrium period of the excited SDS at $H_0 = 0$ satisfies the equation $m = S dT_{eff}/dN$. Based on Eqs. (4), (8), (9), (11) and (13) it is written as

$$P(h) - R(h)\ln\left(a\frac{R(h)}{\tilde{w}(h)}\right) + hR(h)\frac{\tanh(h/2)}{\tilde{w}^2(h)}$$

$$- 4hR(h)\ln\left(ea\frac{R(h)}{\tilde{w}(h)}\right)\tanh\left(\frac{h}{2}\right)\frac{\tilde{w}^2 - \tilde{w}^2(h)}{[\tilde{w}^2 - \tilde{w}^2(h)]^2 + (b\tilde{w})^2} = 0, \tag{14}$$

where

$$P(h) = \frac{l}{h} - \frac{4}{ph^2}\sum_{n=1}^{\infty}\frac{1-(1+h(2n-1))e^{-h(2n-1)}}{(2n-1)^3}, \tag{15}$$

$$R(h) = p\frac{(\tilde{H}\tilde{w})^2}{[\tilde{w}^2 - \tilde{w}^2(h)]^2 + (b\tilde{w})^2}, \tag{16}$$

$\tilde{w} = w/\Omega$, $\tilde{w}(h) = 2\ln^{1/2}(\cosh(h/2))$, $a = pe(2mh)^{1/2}MhL_y/\hbar$, $b = 1/m\Omega$, and $\tilde{H} = H/4pM$. Note that if the oscillating magnetic field is absent then Eq. (14) is reduced to the well-known equation $P(h) = 0$ [10].

We solved Eq. (14) numerically and calculated the dependences of **h** on $\tilde{w}$ and $\tilde{H}$. The dependence of **h** on $\tilde{w}$ shown in Fig. 2 has qualitatively the same form as the experimental curves [6]. As for the dependence of **h** on $\tilde{H}$, the behaviour of calculated and experimental curves at large values of $\tilde{H}$ is different (see Fig. 3). This fact is explained in the following way. The effective temperature (4) was obtained at the condition $\max|x_n| \leq r_c$. For the considered bubble film $r_c \sim 10^{-5}$ cm [12] and Eq. 1 yields $\max|x_n| \sim 10^{-4}\tilde{H}$ cm. From this it follows that the numerical results are correct if $\tilde{H} < 0.1$. Note also that at $H > H_f$ the evaluation of the effective temperature obtained on the basis of the system of stochastic differential equations for $x_n$ yields $T_{eff} \sim H^{-1}$. This result agrees with experimental dependence of **h** on $\tilde{H}$ at large $\tilde{H}$.

## 5. Conclusion

It is shown that the SDS excited by an oscillating magnetic field is a thermodynamic system which is characterized by very high effective temperature. The thermodynamic functions of such domain structure are found, and dependences of its period on the frequency and amplitude of an oscillating magnetic field are calculated. It is established that the peculiarities of these dependences, their nonmonotony in particular, are caused by the entropy term in the free energy of SDS.

## References

[1] T.R. Haller and J.J. Kramer, J. Appl. Phys. 41 (1970) 1034.
[2] Yu.S. Shur, V.A. Zaikova and E.B. Khan, Fiz. Met. Metallov. 29 (1970) 770.
[3] V.K. Vlasko-Vlasov, L.M. Dedukh, V.I. Nikitenko and L.S. Uspenskaya, Fiz. Tverd. Tela 24 (1982) 1255.
[4] V.K. Vlasko-Vlasov and A.F. Khapikov, Zh. Tekh. Fiz. 59 (1989) 91.
[5] V.A. Ignatchenko, P.D. Kim, E.Yu. Mironov and D.C. Khvan, Zh. Eksp. Teor. Fiz. 98 (1990) 593.
[6] V.K. Vlasko-Vlasov and L.S. Uspenskaya, Zh. Eksp. Teor. Fiz. 90 (1986) 1755.
[7] V.M. Eleonskii, A.K. Zvezdin and V.G. Red'ko, Fiz. Met. Metallov. 43 (1977) 7.
[8] E.V. Liverz, Fiz. Tverd. Tela 24 (1982) 3526.
[9] V.G. Bar'yakhtar, Yu.I. Gorobets and S.I. Denisov, Ukr. Fiz. Zh. 28 (1983) 436.
[10] C. Kooy and U. Enz, Philips Res. Rep. 15 (1960) 7.
[11] R. Balescu, Equilibrium and nonequilibrium statistical mechanics (Wiley, New York, 1975).
[12] A.N. Grigorenko, S.A. Mishin and E.G. Rudashevskii, Fiz. Tverd. Tela 30 (1988) 2948.

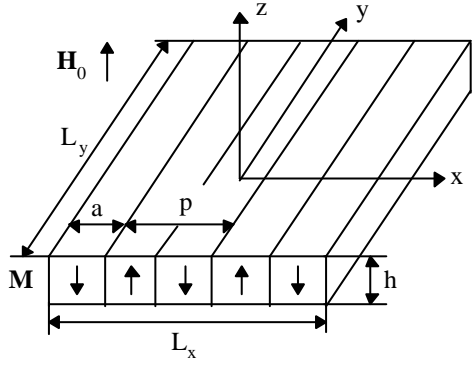

Fig. 1. Schematic view of a bubble film with a strip domain structure.

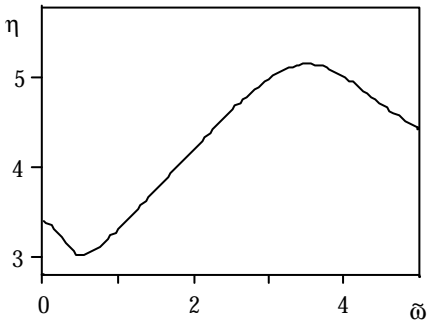

Fig. 2. Dependence of the dimensionless period SDS $h$ on the reduced frequency $\tilde{w}$ of an oscillating magnetic field. The calculation is carried out for a bubble film which has the parameters $h = 10^{-3}$ cm, $L_y = 10^{-2}$ cm, $l = 10^{-4}$ cm, $m = 10^{-10}$ g cm$^{-2}$, $4p M = 600$ G, $b = 5$, and the reduced amplitude $\tilde{H}$ of an oscillating magnetic field is equal to 0.1.

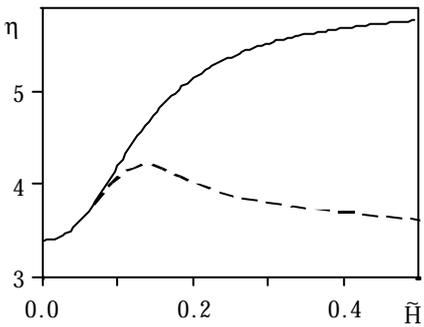

Fig. 3. Calculated (full line) and experimental (dashed line, qualitative view [6]) dependences of the dimensionless period SDS $h$ on the reduced amplitude $\tilde{H}$ of an oscillating magnetic field. The parameters of a bubble film are the same as in the caption to Fig. 2, and the reduced frequency $\tilde{w}$ of an oscillating magnetic field is equal to 2.